\documentclass[preprint]{elsarticle}

\usepackage{lineno,hyperref}
\modulolinenumbers[1]
\usepackage{textcomp}
\journal{arXiv}

\begin{document}

\begin{frontmatter}

\title{Studying the operation of an all-PM Yb-doped fiber laser oscillator at negative and positive net cavity dispersion}


\author[mymainaddress]{Mateusz Pielach\corref{mycorrespondingauthor}}
\cortext[mycorrespondingauthor]{Corresponding author}
\ead{mgpielach@ichf.edu.pl}

\author[mysecondaryaddress]{Jan Szczepanek}

\author[mymainaddress]{Katarzyna Krupa}

\author[mymainaddress]{Yuriy Stepanenko}

\address[mymainaddress]{Institute of Physical Chemistry Polish Academy of Sciences, Kasprzaka 44/52, 01-224 Warsaw, Poland}
\address[mysecondaryaddress]{Fluence sp. z o. o., Kolejowa 5/7, 01-217 Warsaw, Poland}

\begin{abstract}
Chirped fiber Bragg gratings in Yb-doped fiber lasers allow tuning the net cavity dispersion, thus enabling access to pulse dynamics beyond the typical all-normal-dispersion dissipative soliton regime. This Letter demonstrates an ultrafast dispersion-managed all-polarization-maintaining fiber oscillator mode-locked with a nonlinear optical loop mirror. Using a chirped fiber Bragg grating inside the oscillator, it can work at net cavity dispersion in the range from –0.098~$\mathrm{ps^2}$ up to +0.067 $\mathrm{ps^2}$ and above at the wavelength of 1 \textmu m. Depending on the configuration, the system can deliver stable pulses with energies up to 2.3 nJ and pulse durations as short as 98 fs. 
\end{abstract}


\end{frontmatter}


\section{Introduction}

Environmental stability is a key factor distinguishing ultrafast all-polarization-maintaining (PM) fiber lasers from free-space systems~\cite{Hansel, articleShen}. Robustness to vibrations, high temperature, dustiness, or humidity allowed PM-fiber-based light sources to find many applications outside the research laboratories. Such systems are commonly utilized in micromachining \cite{SZYMBORSKI20211496}, imaging \cite{Brinkmann:19, Yang2021}, and medicine \cite{OrringerRaman}. Moreover, the lack of necessity for servicing or aligning optical elements makes fiber lasers more user-friendly and reduces maintenance activities. Consequently, the demand for reliable systems generating ultrashort pulses has significantly grown over the last few years. Therefore, lots of research and industrial efforts optimize state-of-the-art systems towards shorter pulses and larger energies. 

A pulsed regime can be obtained thanks to mode-locking, which in all-PM fiber lasers is usually realized using material or artificial saturable absorbers (SAs). However, material SAs suffer from a low damage threshold, and their saturable absorption capabilities can change over time~\cite{Viskontas}. Hence, artificial SAs are desired in all-PM cavities. The use of the cross-splicing nonlinear polarization evolution (NPE) technique in PM fibers has recently attracted significant research interest \cite{Szczepanek:17, Zhou:18, 9028227, Yu:20}. Nonetheless, most reliable systems are based on a well-developed robust technique incorporating an unequal fiber coupler, which forms a fiber loop that works as a nonlinear amplifying loop mirror (NALM)~\cite{Aguergaray} or a nonlinear optical loop mirror (NOLM) \cite{Szczepanek:15, MYSELF1, BORODKIN2021107353}.

Dispersion management is used to enhance the performance of the mode-locked fiber lasers. In particular, tuning the net cavity dispersion (NCD) influences optical spectra, pulse duration, and energy. Depending on the sign of the NCD, different characteristics of pulsed work can be obtained. All-normal-dispersion (ANDi) fiber lasers working in a dissipative soliton (DS) regime are preferred to obtain the highest pulse energy. On the other hand, to achieve shorter pulses and broader optical output spectra, a dispersion-managed regime is often more desired~\cite{Turitsyn}.

There are a variety of methods capable of changing the dispersion of the cavity. The easiest one is based on two different types of fibers, which are characterized by the opposite signs of the group velocity dispersion (GVD). This technique is common for Er-doped fiber lasers, in which the passive fiber has a different sign of GVD than the active fiber \cite{Sun:18, Laszczych:21}. However, for the exciting wavelength region of around 1 um, standard single-mode PM silica fibers are characterized by a positive GVD. Therefore, most Yb-doped all-fiber lasers possess an ANDi cavity. The most convenient method for introducing negative dispersion to the 1~\textmu m all-PM-fiber cavity, besides employing specialty fibers with GVD tailored by mode dispersion, is to use chirped fiber Bragg gratings (CFBGs).

\begin{figure}[htbp]
\centering
\includegraphics[width=0.99\linewidth]{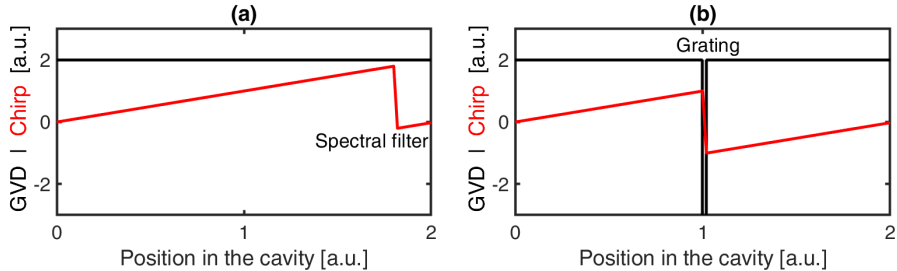}
\centering
\caption[justification=justified]{Schematic dispersion maps for DS (a) and dispersion-managed DS (b). The chirp is marked red, while the black curve indicates GVD. ANDi systems operating in DS regime reduce the chirp only by spectral filtering. In contrast, in the dispersion-managed DS setups, the grating introduces the negative GVD and spectral filtering, resulting in a significant drop of the chirp.}
\label{regimes}
\end{figure}

The fabrication method for inscribing the CFBG inside the fiber limits the range of the optical spectrum, which can be reflected \cite{s18072147}. Thus, a CFBG also introduces spectral filtering. Dispersion management and spectral filtering permit a hybrid regime, called dispersion-managed DS \cite{Bale:09}. Dispersion maps presenting the difference between DS in ANDi systems and dispersion-managed DS regime are schematically depicted in Fig. \ref{regimes}. Moreover, due to limited reflectivity, the grating simultaneously acts as a coupler. Thanks to their simple design and a variety of functions, CFBGs are intensively researched towards their usage in Yb-doped fiber lasers \cite{Zhang:13, YANG2019103008}.
The first works towards incorporating a CFBG in a Yb-doped fiber laser cavity used non-PM (standard step-index single-mode) fibers. The behavior of lasers was presented for both positive and negative NCD \cite{KATZ2007156, Ortac:07}.
Currently, many efforts are being put into implementing CFBGs into all-PM oscillators. An ultrafast setup mode-locked with a semiconductor saturable absorber mirror working at near-zero NCD was shown by Yan \textit{et al.} \cite{Yan:20}. Recently, oscillators mode-locked via artificial SAs have also been investigated. Yu \textit{et al.} built an all-PM Yb-doped oscillator mode-locked using cross-splicing NPE method \cite{Yu:20}. They characterized the system for both positive and negative NCD. However, under negative NCD, their laser was working only in a noise-like regime. NALM-based all-PM fiber laser emitting 88 fs pulses of energy of 0.94 nJ operating at the wavelength of 1 um with a fixed NCD of –0.01 $\mathrm{ps^2}$ was demonstrated by Ma \textit{et al.} \cite{Ma:21}. 

This paper aims to present dispersion management using a CFBG in an all-PM Yb-doped fiber laser mode-locked via NOLM instead. We experimentally investigated the oscillator's characteristics under different values of NCD between –0.098 $\mathrm{ps^2}$ and +0.067 $\mathrm{ps^2}$. Unlike previous all-PM systems, our oscillator can operate in a dispersion-managed DS regime under a broader range of NCD, involving both positive and negative values. Moreover, we present the configuration delivering the highest energy for the negative NCD compared to previously mentioned all-PM setups incorporating CFBGs.

\section{Experimental setup}

\begin{figure}[htbp]
\centering
\includegraphics[width=0.99\linewidth]{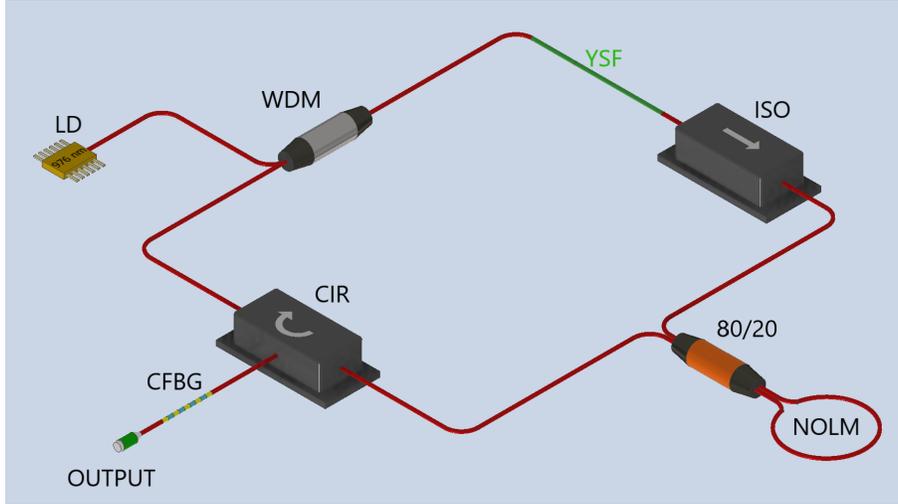}
\centering
\caption[justification=justified]{Scheme of the laser. LD - laser diode; WDM - wavelength division multiplexer; YSF - ytterbium-doped fiber; ISO - optical isolator; NOLM - nonlinear optical loop mirror; CIR - optical circulator; CFBG - chirped fiber Bragg grating.}
\label{scheme}
\end{figure}

Fig. \ref{scheme} shows the scheme of the oscillator. The whole cavity consists of single-mode, single-clad, PM PANDA optical fibers and fiberized components. PM980 fiber was used as a passive medium, while a 0.55 m long piece of PM-YSF-HI was utilized as a gain medium. The group velocity dispersion of both fibers was equal to 0.023 $\mathrm{ps^2/m}$ at 1030 nm. The core of Yb-doped fiber was pumped at 976 nm by a continuous-wave single-mode semiconductor laser diode (3SP-1999CHP) through a wavelength division multiplexer (WDM). A dedicated driver controlled the laser diode, and the maximum pumping power could not exceed 900 mW. An optical isolator (ISO) guaranteed pulse propagation in a clockwise direction. The NOLM loop based on a 2 x 2 type 80/20 coupler served as an artificial SA enabling mode-locking operation. The CFBG was introduced to the cavity through an optical circulator (CIR). The group delay dispersion of the grating was –0.237 $\mathrm{ps^2}$ at 1030 nm. Furthermore, its reflection spectrum possesses a Gaussian shape with a 3~dB spectral bandwidth of 10.4 nm and 32\% peak reflectivity at 1033 nm. Thus, CFBG acts simultaneously as a dispersion managing element, a bandpass filter, and an output coupler. 

In all experiments, output spectra were measured by an optical spectrum analyzer Yokogawa AQ6370C with 0.02 nm resolution. The single-pulse operation was verified with a 1.2 GHz InGaAs photodiode Thorlabs DET01CFC and the radio frequency (RF) spectrum analyzer Agilent Technologies E4443A. We measured RF spectra with 100 Hz resolution bandwidth and 1 kHz span. Thorlabs PM16-405 power sensor was used to measure the average output power. Besides, we performed temporal characteristics of the output pulses using a second harmonic, background-free autocorrelator. Furthermore, we compressed chirped pulses by a parallel diffraction gratings compressor (800 lines/mm). SPIDER technique was used to measure the spectral phase and retrieve the duration of the compressed pulses.

\section{Experimental results and discussion}

We investigated the characteristics of the oscillator for various NCD, aiming to decrease this parameter towards negative values. We tuned the NCD by changing the lengths of particular fibers inside the cavity, which simultaneously scaled the pulse repetition rate. Thanks to spectral filtering introduced by the CFBG, we could obtain a stable single-pulse Raman-free dispersion-managed DS regime for the positive NCD between +0.067 $\mathrm{ps^2}$ and +0.024~$\mathrm{ps^2}$. Note, however, that the oscillator could also operate at larger positive NCD. For the anomalous dispersion, we achieved a dispersion-managed DS regime for NCD values between –0.016 $\mathrm{ps^2}$ and –0.098 $\mathrm{ps^2}$. Nonetheless, for the near-zero NCD between –0.016 $\mathrm{ps^2}$ and +0.024 $\mathrm{ps^2}$, the operation of the laser was unstable.

The behavior of the ANDi fiber lasers was broadly investigated in the literature \cite{PhysRevA.77.023814}, including all-PM Yb-doped NOLM systems \cite{Szczepanek:15, MYSELF1, BORODKIN2021107353}. In this Letter, we prove that a setup incorporating CFBG possesses characteristics similar to ANDi systems for a positive NCD. Since the main purpose of utilizing CFBGs is obtaining near-zero or negative NCD, we characterize one configuration of the smallest positive NCD and compare its characteristics with a negative NCD setup. Particular attention will be devoted to anomalous dispersion configurations.

\begin{figure}[t]
\centering
\includegraphics[width=0.99\linewidth]{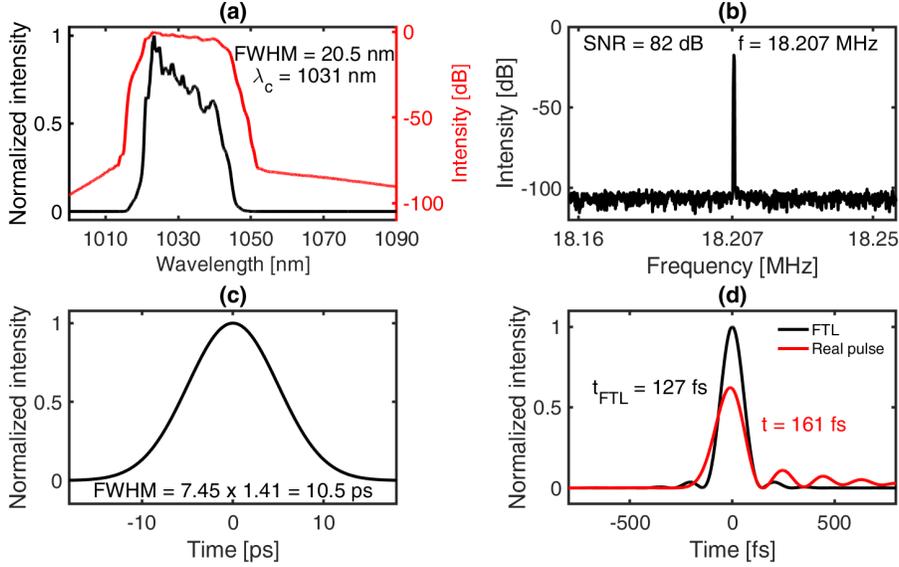}
\centering
\caption[justification=justified]{Spectral and temporal characteristics for the NCD of +0.024 $\mathrm{ps^2}$. (a) Typical steep edges are seen in the optical output spectra in linear (black) and logarithmic (red) scales. (b) A single peak in the radio frequency spectrum proves the stable single-pulse regime. (c) The autocorrelation trace of the output 7.45 ps pulses. (d) The pulses were compressed down to 161 fs (red) while the Fourier transform limit (FTL) is 127 fs (black).}
\label{K175}
\end{figure}

Figure \ref{K175} presents the characteristics of the oscillator operating under NCD of +0.024 $\mathrm{ps^2}$ for the pumping power of 110 mW, taking into account optical losses introduced by the WDM. Optical spectra, characterized by the central wavelength ($\lambda_{\mathrm{c}}$) of 1031 nm, feature a similar shape as ANDi systems. Sharp peaks typical for the DS regime can be observed at the edges of the spectrum. The laser operating at a repetition rate of 18.207~MHz generates ultrashort 7.45 ps pulses of the average output power of 12.6~mW, corresponding to the pulse energy of 0.7 nJ. The maximum pulse energy was limited by the formation of the continuous wave component. Furthermore, the pulses can be compressed down to 161 fs with the temporal Strehl ratio of 0.62. Indeed, the pulse energy is reduced in comparison to previously mentioned ANDi setups \cite{Szczepanek:15, MYSELF1, BORODKIN2021107353}. The limit of the maximum pulse energy for this configuration is most likely caused by the NCD value being relatively close to zero. Nevertheless, dispersion management allowed broadening the output spectra to the full width at half-maximum (FWHM) exceeding 20 nm, reducing the compressed pulse duration.

\begin{figure}[htbp]
\centering
\includegraphics[width=0.99\linewidth]{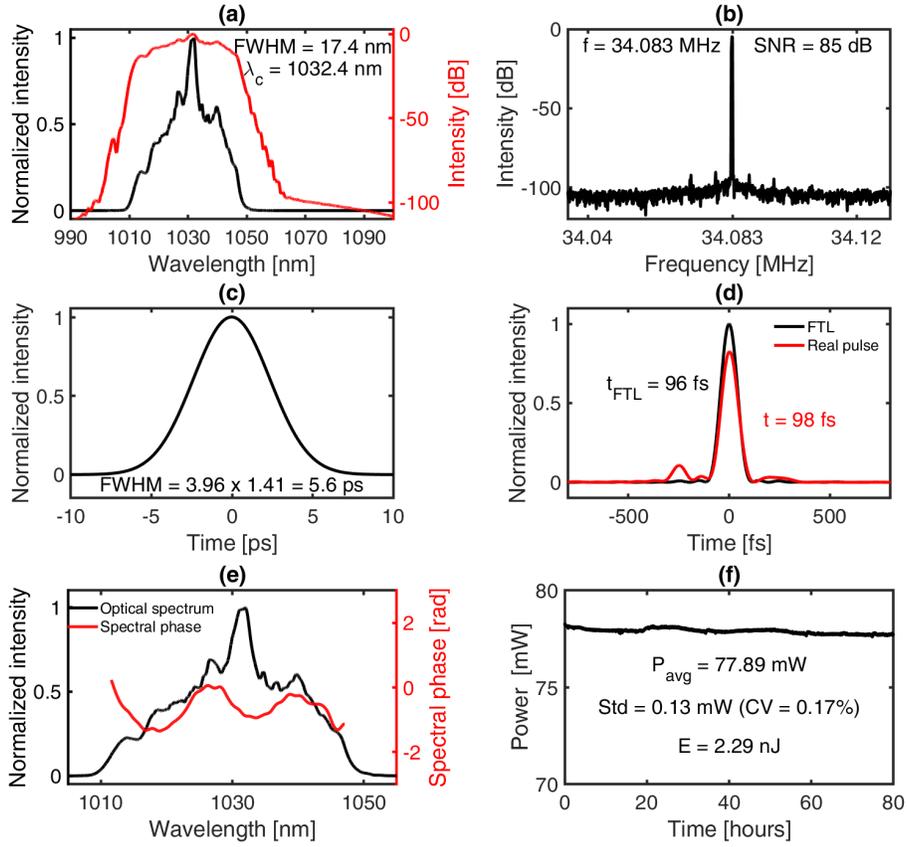}
\centering
\caption[justification=justified]{Spectral and temporal characteristics for the NCD of -0.098 $\mathrm{ps^2}$. (a) The optical output spectra in linear (black) and logarithmic (red) scales lack the typical sharp peaks and are arising more smoothly compared to ANDi systems. (b) A single peak proves operation at a repetition rate of 34.083 MHz in the radio frequency spectrum. (c) The duration of the chirped pulse indicated by the autocorrelation trace is 3.96 ps. (d) Output pulses can be compressed down to 98 fs (red) while the FTL is 96 fs (black). (e) Optical spectrum (black) and spectral phase (red), presenting minor higher-order components. (f)~Average output power as a function of time, proving long-term stability.}
\label{K171}
\end{figure}

When decreasing the lengths of the fibers in the cavity, we run the laser under negative NCD in a dispersion-managed DS regime. Spatial and temporal characteristics for the net anomalous dispersion of -0.098 $\mathrm{ps^2}$ are demonstrated in Figure \ref{K171}. Undoubtedly, a change in the optical spectrum is observed, as it is arising more smoothly. The RF spectrum, containing a single peak at 34.083 MHz with a noise band level below 85 dB, indicates that negative NCD did not impact the stability of the single-pulsed operation. Even though the NCD was negative, the output pulses remained positively chirped. Nevertheless, the duration of the chirped pulse was reduced to 3.96 ps. Moreover, we achieved pulses compressible almost to the Fourier transform limit (FTL) of 96 fs. Despite the presence of minor higher-order components in the spectral phase, we dechirped the pulses down to 98 fs with an improved temporal Strehl ratio of 0.83. 
Note that although the FWHM of the output spectrum slightly decreased when compared to the reported above positive NCD configuration, the pulses could be compressed to shorter durations.
At the pumping power of 260 mW, the oscillator's average output power ($\mathrm{P_{avg}}$) was equal to 77.9 mW, corresponding to the pulse energy of 2.3 nJ. Furthermore, the oscillator features excellent power stability over time, proved by the standard deviation (Std) of 0.13 mW, leading to the coefficient of variance (CV) below 0.2\%. On the other hand, the presented configuration of a negative NCD is not self-starting, meaning that increasing the pumping power is not sufficient to initialize the mode-locking operation of the laser. One has to introduce a mechanical disturbance, such as tapping the optical fibers. However, any further mechanical perturbations do not affect the laser's performance once the oscillator starts working in a pulsed regime.

\begin{figure}[htbp]
\centering
\includegraphics[width=0.99\linewidth]{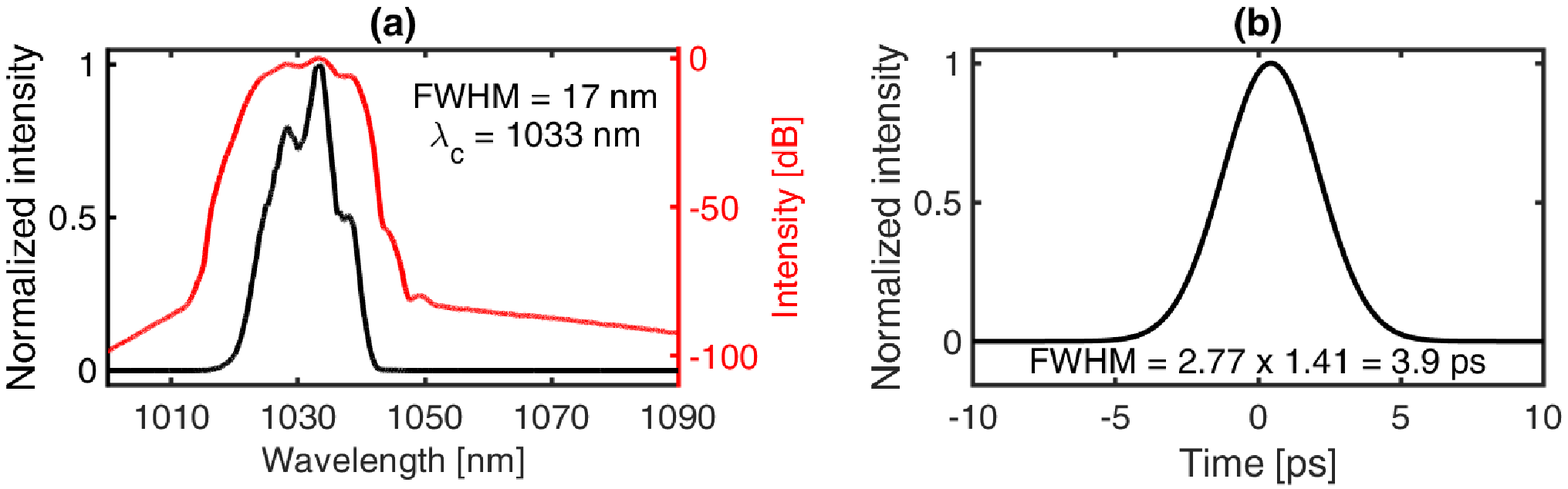}
\centering
\caption[justification=justified]{Spectral and temporal characteristics for the NCD~of -0.051 $\mathrm{ps^2}$. (a) The optical output spectra in linear (black) and logarithmic (red) scales are characterized by weaker modulation. (b) The duration of the chirped pulse decreased to 2.77 ps.}
\label{K172}
\end{figure}

After insightful analysis, we observed that the further from zero is the NCD, the higher the maximum pulse energies. While the closer to zero is the NCD, the lower is the optimal pumping power at which the operation of the oscillator is stable. The laser running under NCD of -0.051 $\mathrm{ps^2}$, corresponding to the repetition rate of 25.537 MHz, could operate at the pumping power of 100 mW. For this case, the pulse energy did not exceed 0.5~nJ.
Fig. \ref{K172} presents the characteristics of the laser operating under NCD of -0.051 $\mathrm{ps^2}$. Due to weaker self-phase modulation, the optical spectrum is less ragged compared to the oscillator's operation at NCD of -0.098 $\mathrm{ps^2}$. However, the spectrum is narrower, leading to the increase of the FTL pulse duration. On the other hand, the uncompressed pulse duration decreased to 2.77~ps. 

\begin{figure}[!htbp]
\centering
\includegraphics[width=0.99\linewidth]{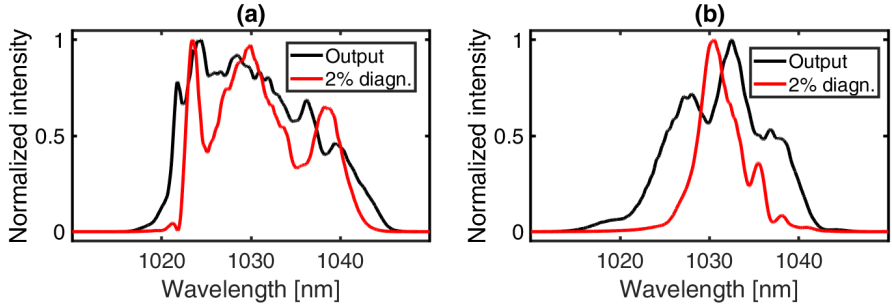}
\centering
\caption[justification=justified]{Spectral characteristics in two different places of the cavity: output (black) and 2\% diagnostic port before the NOLM loop (red) for
the NCD of +0.045 $\mathrm{ps^2}$ (a) and -0.047 $\mathrm{ps^2}$ (b). Significant spectral broadening induced by the self-phase modulation effect can be seen for the negative dispersion.}
\label{K98}
\end{figure}

The operation of the oscillator under a negative NCD differs from the characteristics in a net positive dispersion, especially in terms of the pulse durations and the shape of the optical output spectra. To provide further insight into the reported dynamics of the dispersion-managed DS regime, we performed a preliminary study towards revealing the differences in pulse evolution inside the cavity. We spliced an additional diagnostic 98/2 fiber coupler between the isolator and the NOLM loop, so that 2\% of the power is extracted from the cavity. We measured the spectra in two different positions in the cavity for two configurations with NCD values placed almost symmetrically near zero. Fig. \ref{K98}(a) compares obtained spectra for the setup running under positive NCD of +0.045 $\mathrm{ps^2}$, that corresponds to the repetition rate of 16.8 MHz. The spectra for the net anomalous dispersion of -0.047 $\mathrm{ps^2}$ running at 25 MHz are shown in Fig. \ref{K98}(b) instead. The pumping power for both results was equal to 100 mW. Indeed, the evolution of the spectra can be observed for both cases. However, for the negative NCD, the spectral broadening is more significant. We believe that this is caused by a smaller chirp in the cavity, leading to higher peak power, thus a more substantial impact of the self-phase modulation effect, which broadens the spectra. Nevertheless, further intracavity measurements should be carried out to fully understand the breathing dynamics of the pulses.

\section{Conclusion}

To sum up, we have presented the first, to the best of our knowledge, dispersion-managed Yb-doped all-PM fiber laser oscillator mode-locked via\break NOLM. Unlike previous ANDi NOLM-based all-PM Yb-doped setups, our oscillator can also operate under negative NCD. The oscillator's operation under positive dispersion features similar spectral characteristics as ANDi systems \cite{PhysRevA.77.023814, Szczepanek:15}. Thanks to the CFBG, operating under negative NCD significantly reduced the pulse duration and broadened the spectra while sustaining the stability of the pulsed operation. Thus, the dispersion-managed DS regime opens up a way towards generating shorter pulses than ANDi systems. On the other hand, for negative NCD, the pulse is still positively chirped due to the position of the output coupler. Note that the smoother shape of the output spectra results in the reduction of the FTL pulse duration. Thus, after external compression, it is possible to obtain shorter pulses. We believe that further spectral broadening and optimizing the NCD as well as the position of the output coupler may result in obtaining even shorter pulses. However, there is a trade-off between pulse duration and energy. We reckon that the system can also be optimized towards more considerable energies, as we did not see stimulated Raman scattering at any point \cite{Szczepanek:s} and reaching higher energies were limited by the formation of the continuous wave component. Nevertheless, the obtained energy of 2.3 nJ is more than two times larger than the highest energy reported so far in a class of dispersion-managed Yb-doped all-PM systems for similar pulse durations in the sub-100 fs range \cite{Ma:21}. To conclude, simple design, environmental robustness, short pulse duration, and nJ-level pulse energy prove the potential of dispersion-managed NOLM-based oscillator, especially in applications where shorter pulses or broader spectra are desirable.

\section*{Declaration of competing interest} 
The authors declare that they have no known competing financial interests or personal relationships that could have appeared to influence the work reported in this paper.

\section*{Funding}
This work was funded by the Foundation for Polish Science, under research project TEAM-NET No. POIR.04.04.00-00-16ED/18.

\bibliography{bibfile}

\end{document}